\RequirePackage{lineno}
\documentclass[aps,prc,twocolumn, floatfix]{revtex4}
\usepackage{graphicx}
\usepackage{dcolumn}
\usepackage{bm}
\usepackage{mhchem}
\usepackage{color}
\usepackage{multirow}

\begin{document}


\title{Exploring hadron-quark phase transition in heavy-ion collisions using particle emission ratios in heavy and light reaction systems}

\author{Xun Zhu$^{1,2}$}
\author{Gao-Chan Yong$^{1,2}$}
\email[Corresponding author: ]{yonggaochan@impcas.ac.cn}

\affiliation{
$^1$Institute of Modern Physics, Chinese Academy of Sciences, Lanzhou 730000, China\\
$^2$School of Nuclear Science and Technology, University of Chinese Academy of Sciences, Beijing 100049, China
}

\begin{abstract}

Based on the AMPT model, which incorporates both hadronic and quark degrees of freedom, we studied the productions of lambda, kaon, proton, and pion in reaction systems $^{40}$Ca+$^{40}$Ca, $^{48}$Ca+$^{48}$Ca, and $^{197}$Au+$^{197}$Au. It is found that the ratios of identical particle emissions from heavy and light reaction systems, especially the emission ratios of strange particles $\Lambda^{0}$ or K$^{+}$ in heavy and light reaction systems, are highly sensitive to the hadron-quark phase transition in heavy-ion collisions. Detailed explanations and validations of these results are given.

\end{abstract}

\maketitle

%
A few microseconds following the Big Bang, the universe existed in a hot and dense state known as quark-gluon plasma (QGP). This initial state of matter transitioned into hadronic matter as the universe expanded, a process explained by Quantum Chromodynamics (QCD) \cite{ellis2006little}. Investigating the quark-gluon plasma and the QCD phase transition is essential for comprehending the early universe's evolution and the universe's structure. It is well-understood that in hadrons, quarks and gluons are confined due to the phenomenon known as color confinement. Researching these high-energy states and their transformations on Earth poses a significant challenge. Relativistic heavy-ion collision experiments tackle this issue \cite{iancu2012qcd,odyniec2022probing} by accelerating and colliding nuclei at extremely high energies. These collisions generate the necessary conditions for forming quark-gluon plasma by depositing energy in a confined space \cite{aidala2023new}. Observations of QGP have been made at both the Relativistic Heavy-Ion Collider (RHIC) and the Large Hadron Collider (LHC) \cite{heinz2001sps,gyulassy2004qgp,niida2021signatures}. Therefore, Studying relativistic heavy-ion collisions is an effective approach to exploring the formation of quark-gluon plasma and the QCD phase transition.

A key objective in studying the theory of QCD phase transitions is to clearly outline the QCD phase diagram. QCD matter presents different phases depending on variations in temperature ($T$) and baryon chemical potential ($\mu_B$). At lower temperatures or higher baryon chemical potentials, matter mainly consists of hadrons. As the temperature rises or the baryon chemical potential decreases, hadronic matter undergoes transitions that result in the deconfinement of quarks and gluons, forming a quark-gluon plasma. When the baryon chemical potential is close to zero, lattice QCD calculations show that the transition from a hadron gas to a quark-gluon plasma occurs through a smooth crossover \cite{schmidt2006lattice}. In regions of high baryon chemical potential, theoretical calculations suggest a first-order phase transition between these phases \cite{du2024qcd}. There is also a transition point between the first-order and the crossover phase transitions, known as the critical point of the phase transition \cite{stephanov2006qcd}. To precisely chart the phase transition curve, researchers adjust the creation of QGP by controlling initial collision conditions, such as energy, centrality, and collision system \cite{du2024qcd}. These changes in initial conditions determine the type and occurrence of phase transitions. Because of the extremely short-lived nature of QGP and its extreme conditions, direct detection is not possible. Consequently, identifying probes that are sensitive to the QCD phase transition or QGP has become a crucial method for investigating this phenomenon.

Fortunately, several physical quantities have been proposed through both experimental and theoretical efforts, providing new opportunities to explore QCD phase transitions and the properties of QGP. Anisotropic flow \cite{csernai1999third,shen2021studying,gajdovsova2021probing} and jet quenching \cite{baier2002jet} offer crucial insights into the formation and dynamics of QGP. Particle fluctuations \cite{stephanov1999event,ASAKAWA2016299} and light nuclei yield ratios \cite{shuryak2020light} are sensitive indicators for identifying critical points in phase transitions. Furthermore, probes such as heavy flavor meson production \cite{patra2009j}, dilepton production \cite{dusling2008dilepton}, and the generation of strange particles \cite{YONG2023138051} allow researchers to further investigate key aspects of the QCD phase diagram. Most of these probes are in fact based on the distinct properties of specific particles in the same reaction system. In this study, we aim to identify a physical quantity that can be applied to various particles, starting from the scattering mechanisms involved in fundamental hadron and parton transport. Our findings suggest that the ratio of identical particle from heavy and light systems holds promise as a probe for detecting the presence of partons in heavy-ion collisions. Beyond the underlying physical mechanisms, using the particle ratio in heavy and light systems helps reduce systematic errors, thereby enhancing reliability.

Recent studies have advanced hybrid models by dynamically coupling hydrodynamics and transport approaches, enabling core-corona separation and adaptive initialization in high-baryon-density systems \cite{hyd1,hyd4}, while others have extended three-fluid dynamics to event-by-event simulations across GeV-scale energies \cite{hyd3,hyd5}. These developments highlight the importance of multiscale fluid-transport integration for resolving QCD phase transitions and collective dynamics from SPS to BES energies \cite{hyd6,hyd1,hyd2}. While fluid-hybrid methods excel in capturing multiscale dynamics through coupled hydrodynamic and transport descriptions, traditional microscopic transport approaches retain unique strengths in resolving partonic cascades at lower energies, thereby synergistically addressing the full energy landscape of heavy-ion collisions.

%
In the early 2000s, the AMPT (a multiphase transport) model was developed \cite{lin2005multiphase} to analyze results from quark-gluon plasma experiments conducted at RHIC and LHC, using the framework of non-equilibrium many-body dynamics. This model includes two variants: the default version and the string melting (SM) version. In the present study, we employed the SM version of the AMPT model, which starts by initializing the colliding nucleons as a combination of hard minijets (above 10 GeV center of mass energy) and soft excited strings, eventually fragmenting into partons. The ZPC (Zhang's Parton Cascade) model describes two-body elastic scatterings among these partons. After these scatterings, the partons are recombined into hadrons via the quark coalescence model. Finally, the interactions among the hadrons are depicted using a hadronic cascade process based on the ART (A Relativistic Transport) model. For more information, please refer to Ref. \cite{lin2005multiphase}. To describe the nuclear collision process in the absence of parton degrees of freedom, a pure hadron cascade version of the AMPT model (AMPT-HC) has been developed to describe nuclear collisions without considering parton degrees of freedom \cite{yong2021double,YONG2023138051}. In this version, the nucleon density distribution of the colliding nuclei is constructed using a Woods-Saxon profile, with the local Thomas-Fermi approximation applied to ascertain the positions and momenta of the nucleons. Once initialized, the nucleons directly engage in a hadron cascade process using an extended ART model. The primary difference between the AMPT-HC mode and the AMPT-SM mode is the absence or presence of parton degrees of freedom.

Since the AMPT model, as previously introduced, does not account for inelastic parton-parton scattering in its partonic scattering component, and considering the complexity of incorporating inelastic parton scattering, we do not intend to introduce it into the AMPT model. To justify the exclusion of inelastic parton scattering in the AMPT model and to demonstrate that hadronic rescattering leads to an increase in hadron yields, we introduce the following PACIAE model, as it is particularly well-suited for addressing these two issues. In 2023, An-Ke Lei and colleagues introduced the PACIAE 3.0 model, an advancement from PACIAE 2.2 and PYTHIA 6.428 \cite{lei2023introduction}. This Monte Carlo model is designed for simulating nuclear reactions and employs three frameworks: A ($\sqrt{s_{NN}}\leq3\,\mathrm{GeV}$), B, and C to represent various reaction processes. A key attribute of PACIAE 3.0 is its conceptualization of nuclear reactions as an accumulation of hadron-hadron (hh) collisions. At the start of a reaction, nucleon positions within the nucleus are established using the Woods-Saxon distribution, paired with a uniform distribution over the 4$\pi$ solid angle, thus forming the initial particle list. The present analysis concentrates on the B and C frameworks, with detailed explanations of these two frameworks and key highlights provided below. The PACIAE-C framework is comparable to the AMPT-SM mode, as both simulate parton interactions, hadron interactions, and the phase transitions between them. Although both models share an overall structural resemblance, there are significant differences in their execution. In the parton scattering processes, PACIAE-C includes both elastic and inelastic parton scatterings, whereas AMPT only accounts for elastic scatterings of partons. The PACIAE-B framework can be understood through a two-step process. Initially, each hh collision via PYTHIA generates a hadronic state composed of numerous hadrons and their four-dimensional momenta. Subsequently, the hadrons engage in hadronic rescatterings, resulting in the hadronic final state. Notably, in the model employed for this study, nucleons created by each PYTHIA collision are restricted to a single collision opportunity. For hadron rescatterings, newly created particles are allowed to undergo multiple rescatterings.

%
\begin{figure}[t!]
\centering
\includegraphics[width=0.48\textwidth]{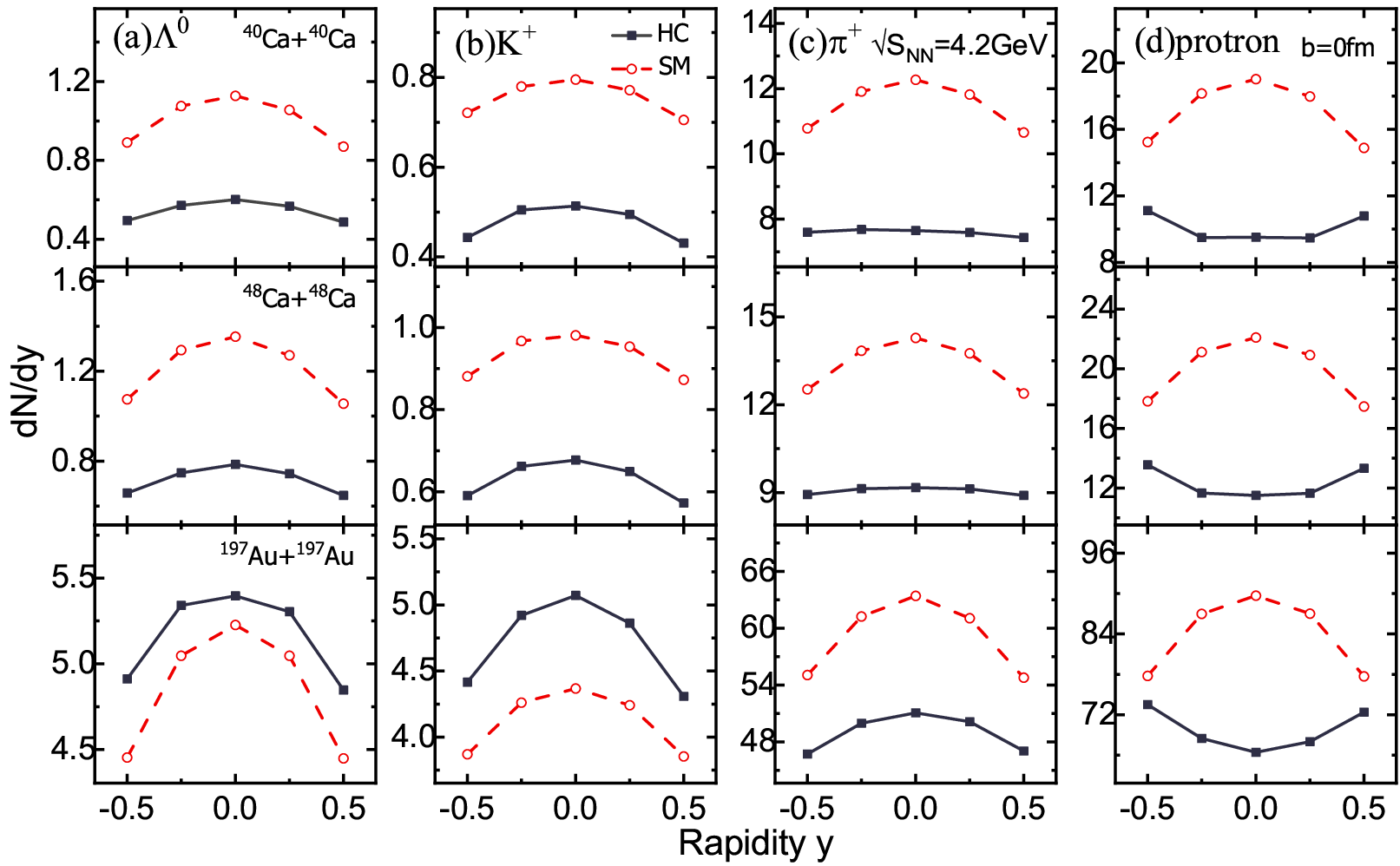}
\caption{Rapidity distributions of particles produced in central nucleus-nucleus collisions at $\sqrt{s_{NN}}=4.2  \,\mathrm{GeV}$, given by the AMPT model. Panels (a) present rapidity distributions of $\Lambda^{0}$ for  $^{40}$Ca+$^{40}$Ca, $^{48}$Ca+$^{48}$Ca, and $^{197}$Au+$^{197}$Au reaction systems (from top to bottom) and panels (b), (c) and (d) for K$^{+}$, $\pi^{+}$ and proton, respectively. The red dashed lines and open circles represent the results obtained from the AMPT-SM mode, while the black solid lines and solid squares denote the results from the AMPT-HC mode.} \label{caau}
\vspace{0.25cm}
\end{figure}
Figure~\ref{caau} presents the rapidity distributions for various particles ($\Lambda^0$, K$^+$, $\pi^+$, and proton) produced in different collision systems (${^{40}\mathrm{Ca}}$+${^{40}\mathrm{Ca}}$, ${^{48}\mathrm{Ca}}$+${^{48}\mathrm{Ca}}$, and ${^{197}\mathrm{Au}}$+${^{197}\mathrm{Au}}$) as computed using the AMPT-HC and AMPT-SM modes. Generally, except for the $\Lambda^0$ and K$^+$ particles in the ${^{197}\mathrm{Au}}$+${^{197}\mathrm{Au}}$ system, the yields from the SM mode surpass those from the HC mode. Nonetheless, for $\Lambda^0$ and K$^+$ particles, as the collision system scales up from ${^{40}\mathrm{Ca}}$+${^{40}\mathrm{Ca}}$ to ${^{48}\mathrm{Ca}}$+${^{48}\mathrm{Ca}}$ and subsequently to ${^{197}\mathrm{Au}}$+${^{197}\mathrm{Au}}$, the yields from the AMPT-HC mode increase relative to those from the AMPT-SM mode, eventually exceeding them. For $\pi^+$ and proton, although the yields from the HC mode do not overtake those from the SM mode as the system size expands, the disparity between the two modes decreases progressively. For instance, at mid-rapidity ($y = 0$), the $\pi^+$ yields using the HC mode constitute 49\%, 52\%, and 74\% of those obtained with the SM mode as the system evolves from ${^{40}\mathrm{Ca}}$+${^{40}\mathrm{Ca}}$ to ${^{48}\mathrm{Ca}}$+${^{48}\mathrm{Ca}}$, and ultimate to ${^{197}\mathrm{Au}}$+${^{197}\mathrm{Au}}$. Putting aside the uncertainties in the particle production mechanisms of different AMPT modes, we hypothesize that this phenomenon arises due to the increased number of hadronic multiple collisions in the HC mode as the collision system grows larger. The HC mode implements a pure hadronic cascade process, and with the increase in hadron numbers in larger systems, the likelihood of multiple hadronic scatterings elevates. This enhancement leads to a higher probability of inelastic collisions for each hadron, which results in an increased number of final-state hadrons. Conversely, the SM mode does not incorporate partonic inelastic scatterings due to their negligible cross sections, which would otherwise accumulate higher partonic collision energy and produce more partons. Since partonic elastic scatterings do not result in an increase in the number of partons, the yield of hadrons, formed through coalescence, remains consistent. Thus, the final-state hadron yield does not experience growth due to multiple partonic scatterings in the SM mode. Therefore, as the system size expands, the growth rate of the final-state hadron yield in the HC mode surpasses that in the SM mode. Although in the ${^{40}\mathrm{Ca}}$ system, the yield of final-state hadrons in the HC mode is lower than that in the SM mode, the accelerated growth in the HC mode with increasing nuclear size leads to a crossover at a specific nuclear size. Further studies have been conducted to validate this hypothesis.

\begin{figure}[t!]
\centering
\includegraphics[width=0.48\textwidth]{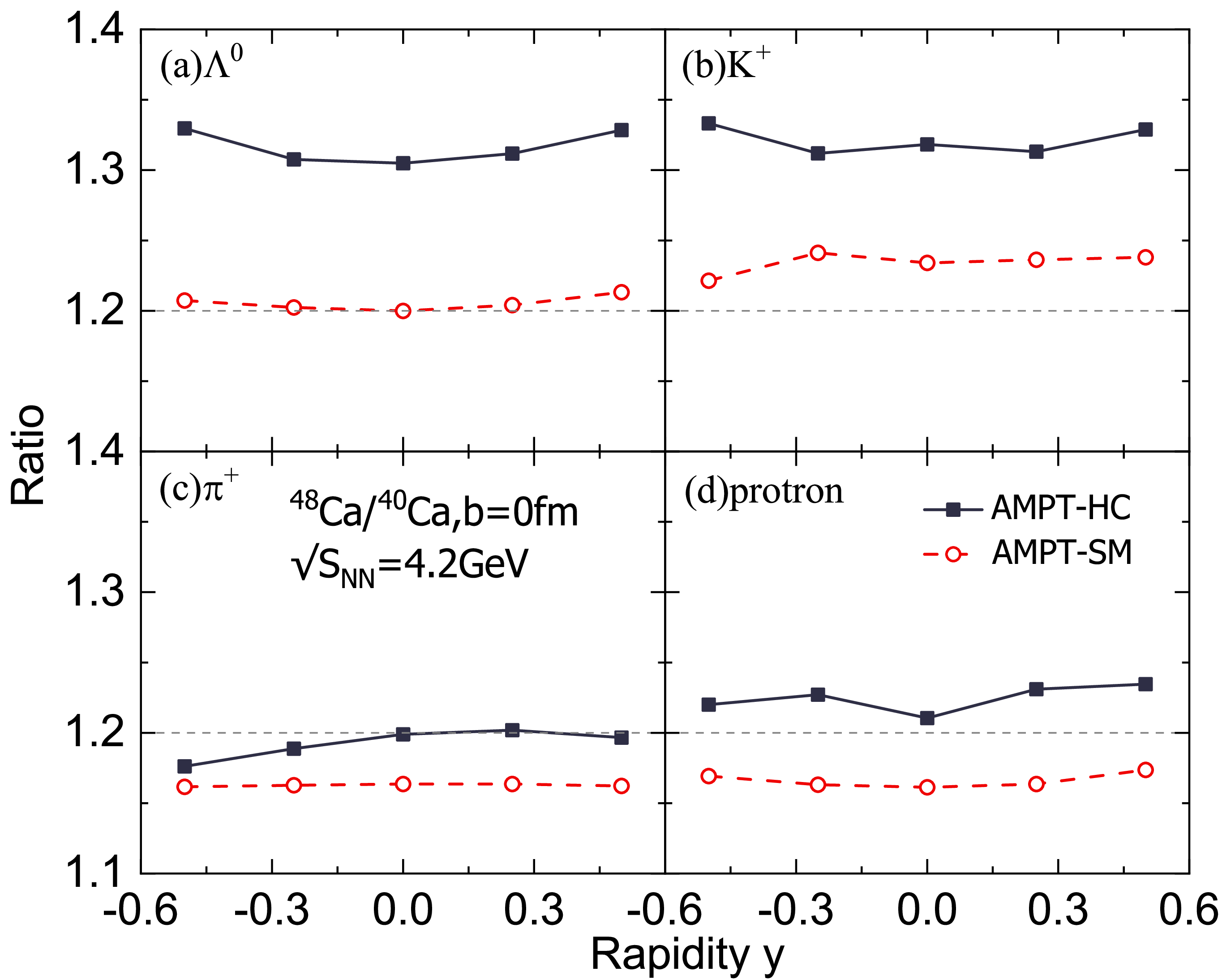}
\caption{Ratios of particle productions as a function of rapidity in the central ${}^{48}\mathrm{Ca}$+${}^{48}\mathrm{Ca}$ and ${}^{40}\mathrm{Ca}$+${}^{40}\mathrm{Ca}$ collisions at $\sqrt{s_{NN}}=4.2  \,\mathrm{GeV}$. Panels (a), (b), (c), and (d) show distributions of $\Lambda^0$, K$^+$, $\pi^+$, and proton, respectively. The dashed line represents the ratio of the total number of nucleons in the two reaction systems.} \label{ca48}
\vspace{0.25cm}
\end{figure}
\begin{figure}[t!]
\centering
\includegraphics[width=0.48\textwidth]{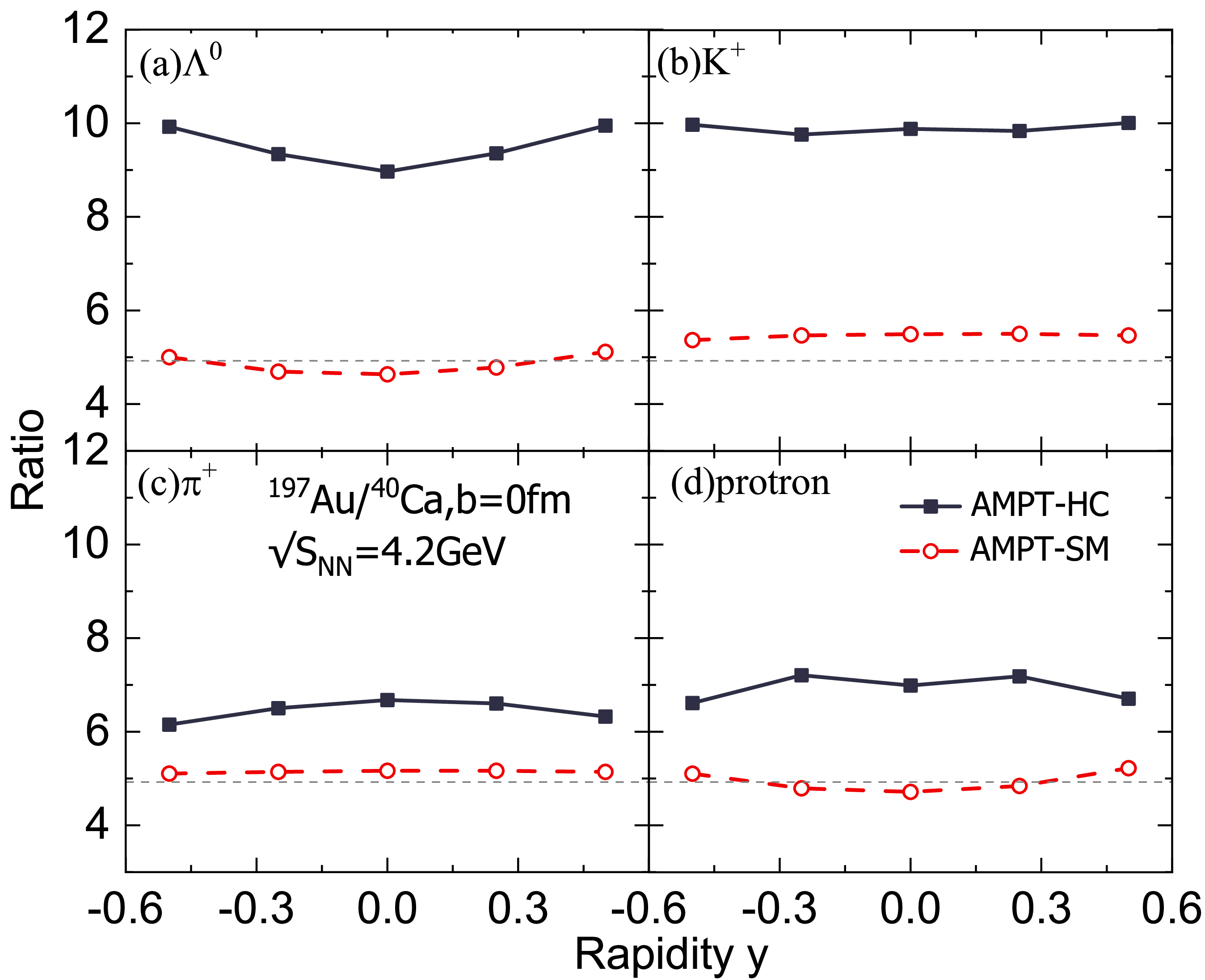}
\caption{Same as Fig.~\ref{ca48}, but for  ${}^{197}\mathrm{Au}$+${}^{197}\mathrm{Au}$ and ${}^{40}\mathrm{Ca}$+${}^{40}\mathrm{Ca}$ systems. The dashed line represents the ratio of the total number of nucleons in the two reaction systems.} \label{au197}
\vspace{0.25cm}
\end{figure}

To more clearly observe the increase in particle production in larger reaction systems using the HC mode compared to the SM mode, we calculated the ratios of particle production between heavy and light reaction systems. Figure~\ref{ca48} displays these ratios as a function of rapidity in the ${}^{48}\mathrm{Ca}$+${}^{48}\mathrm{Ca}$ and ${}^{40}\mathrm{Ca}$+${}^{40}\mathrm{Ca}$ systems. Numerically, the ratio of nucleon numbers between ${}^{48}\mathrm{Ca}$ and ${}^{40}\mathrm{Ca}$ is 1.2. Assuming nucleus-nucleus collisions are a superposition of nucleon-nucleon collisions, the expected particle production ratios in the two systems should be approximately 1.2. It is observed that, although the ratios of particle production generally hover around 1.2, in the HC mode, the ratios of $\Lambda^0$ and K$^+$ productions in the two systems are significantly higher than 1.2. This is attributed to the increased hadron rescatterings in the heavier reaction system under the HC mode, as previously discussed. Additionally, this figure also underscores that for heavy reaction systems, the HC mode can lead to greater particle production when compared to the SM mode. When ${}^{48}\mathrm{Ca}$ is replaced with ${}^{197}\mathrm{Au}$ -- which has a larger nucleon number difference compared to ${}^{40}\mathrm{Ca}$ -- the effects demonstrated in Figure~\ref{ca48} become more pronounced, as illustrated in Figure~\ref{au197}. The nucleon number ratio between ${}^{197}\mathrm{Au}$ and ${}^{40}\mathrm{Ca}$ is 4.925. It is observed that in the SM mode, the ratios for all four particle types are around 4.925; however, with the HC mode, the ratios are consistently greater than 4.925, particularly for the ratios of $\Lambda^0$ and K$^+$ in the two systems. It is worth noting here that the observed asymmetry between positive and negative rapidity in the ratio observables of Figure~\ref{ca48} and Figure~\ref{au197} is a result of insufficient statistical significance.

As discussed earlier, the string melting model for heavy-ion collisions, AMPT-SM, has early-stage parton production proportional to the number of participating nucleons in the reaction system. In the mid-stage, parton scattering, due to the small inelastic scattering cross section of partons, the elastic scattering processes do not increase the number of partons, resulting in no new increase in the number of hadrons formed by parton coalescence in the later stages. Therefore, for the AMPT-SM mode, the final state hadron production is proportional to the reaction system, manifesting as a ratio of particle production numbers in heavy and light reaction systems equivalent to the ratio of nucleon numbers in the reaction systems. Conversely, for the pure hadronic scattering model in heavy-ion collisions, AMPT-HC, hadrons directly undergo multiple scatterings after the initial collision, with a higher probability of multiple scatterings in heavier reaction systems. Multiple scatterings of hadrons in heavier reaction systems lead to a ratio of particle production numbers in heavy and light reaction systems that is greater than the ratio of nucleon numbers in the reaction systems, especially for very heavy compared to very light reaction systems.

Given that the AMPT-SM model, which includes partonic degrees of freedom, and the AMPT-HC model, based purely on hadronic degrees of freedom, have both achieved significant success in high-energy and low-energy heavy-ion collisions respectively \cite{lin2005multiphase,YONG2023138051,yong2024pt,amptrw}, it is noteworthy that in the potential energy region of heavy-ion collisions where a hadron-quark phase transition might occur, both the AMPT-SM and AMPT-HC models are highly sensitive to the ratios of particle yields, particularly the ratios of $\Lambda^0$ and K$^+$ production, between heavy and light reaction systems (this is because the production of strange particles is minimally influenced by final-state interactions, thereby preserving more information from the early stages of the reaction system. In contrast, pion emission, due to multiple scatterings with baryons in the later stages of the reaction, results in significant loss of information from the early stages of the system). Therefore, one can propose a new observable for detecting the occurrence of a hadron-quark phase transition in heavy-ion collisions: the ratio of particle yields, especially the production ratios of $\Lambda^0$ and K$^+$, between heavy and light reaction systems. Should the measured ratios exhibit a closer alignment with the SM scenario, it would indicate the occurrence of a hadron-to-quark phase transition. Conversely, consistency with the HC model would imply the absence of such a transition.
If it is experimentally confirmed that in the small system of Ca+Ca, due to the limited volume of compressed matter produced, a phase transition cannot occur, then one may opt for a larger intermediate-mass reaction system, such as Sn+Sn.

In addition to a high level of sensitivity (pion ratio is approximately 20\%, proton ratio is approximately 40\%, Lambda and kaon ratio is about 100\%), we propose investigating QCD phase transitions by combining heavy and light reaction systems primarily because the comparative analysis of these dual systems more distinctly reveals that, in the absence of a phase transition, hadron multiple scattering escalates exponentially with the size of the reaction system--resulting in an exponential increase in the production of final-state hadrons. This is a signature markedly divergent from that observed during a phase transition. Moreover, using observables from two reaction systems avoids uncertainties in the high-density nuclear matter Equation of State (EoS), theoretical uncertainties in particle production reaction channels and their medium corrections, differences in program algorithm implementations, and different model frameworks, etc., which lead to uncertainties in particle yields. These uncertainties in transport models often accumulate, resulting in greater uncertainties in predicting particle production for a single reaction system than the differences in calculations provided by the AMPT-HC and AMPT-SM modes for a single reaction system in this study. Therefore, we propose that using the ratio of observables from two reaction systems can bypass many theoretical uncertainties in transport model calculations, improving the reliability of interpreting experimental data. Additionally, this analytical approach also helps reduce the interference of various noises in experimental data analysis.

\begin{figure}[tb]
\centering
\includegraphics[width=0.49\textwidth]{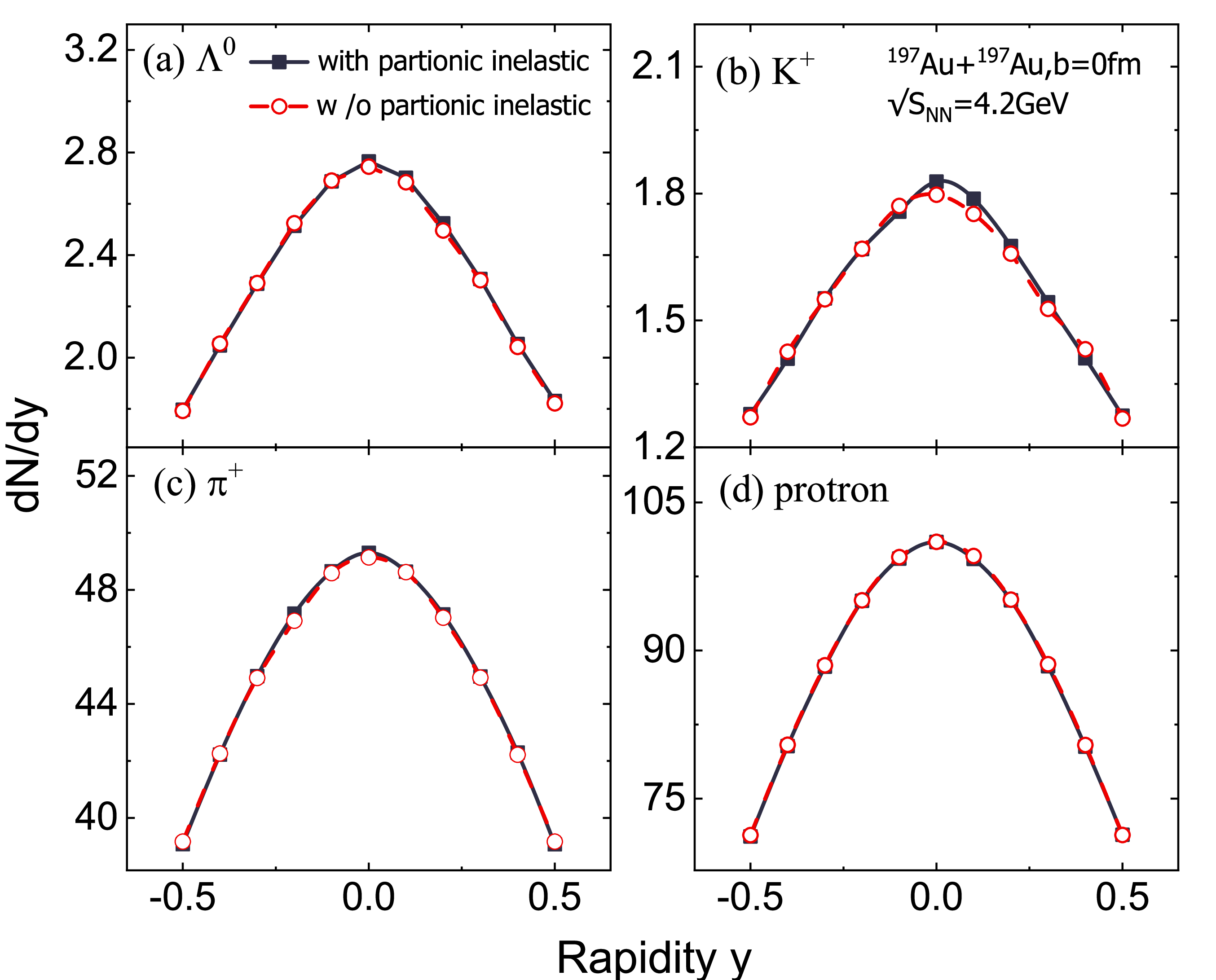}
\caption{Effects of partonic inelastic scatterings on the $\Lambda^0$, K$^+$, $\pi^+$, and proton productions in the central ${}^{197}\mathrm{Au}$+${}^{197}\mathrm{Au}$ collision at $\sqrt{s_{NN}}=4.2  \,\mathrm{GeV}$, given by the PACIAE model. The solid lines and filled squares represent results with partonic inelastic collisions while the dashed lines and open circles represent results without partonic inelastic collisions.} \label{inelastic}
\vspace{0.25cm}
\end{figure}
In order to further verify the reasonableness of omitting partonic inelastic scatterings in the calculations using the AMPT-SM model, it is necessary to check the sensitivity of the previously presented particle productions to the partonic inelastic scatterings. Figure~\ref{inelastic} illustrates the effects of the partonic inelastic scatterings on the $\Lambda^0$, K$^+$, $\pi^+$ and proton productions in the central ${}^{197}\mathrm{Au}$+${}^{197}\mathrm{Au}$ collisions at $\sqrt{s_{NN}}=4.2  \,\mathrm{GeV}$. Here, we employ the PACIAE-C model, which includes partonic inelastic scattering cross sections, for the study. It can be observed that for the productions of $\Lambda^0$, K$^+$, $\pi^+$, as well as proton, the effects of partonic inelastic scatterings can indeed be considered negligible.

\begin{figure}[tb]
\centering
\includegraphics[width=0.49\textwidth]{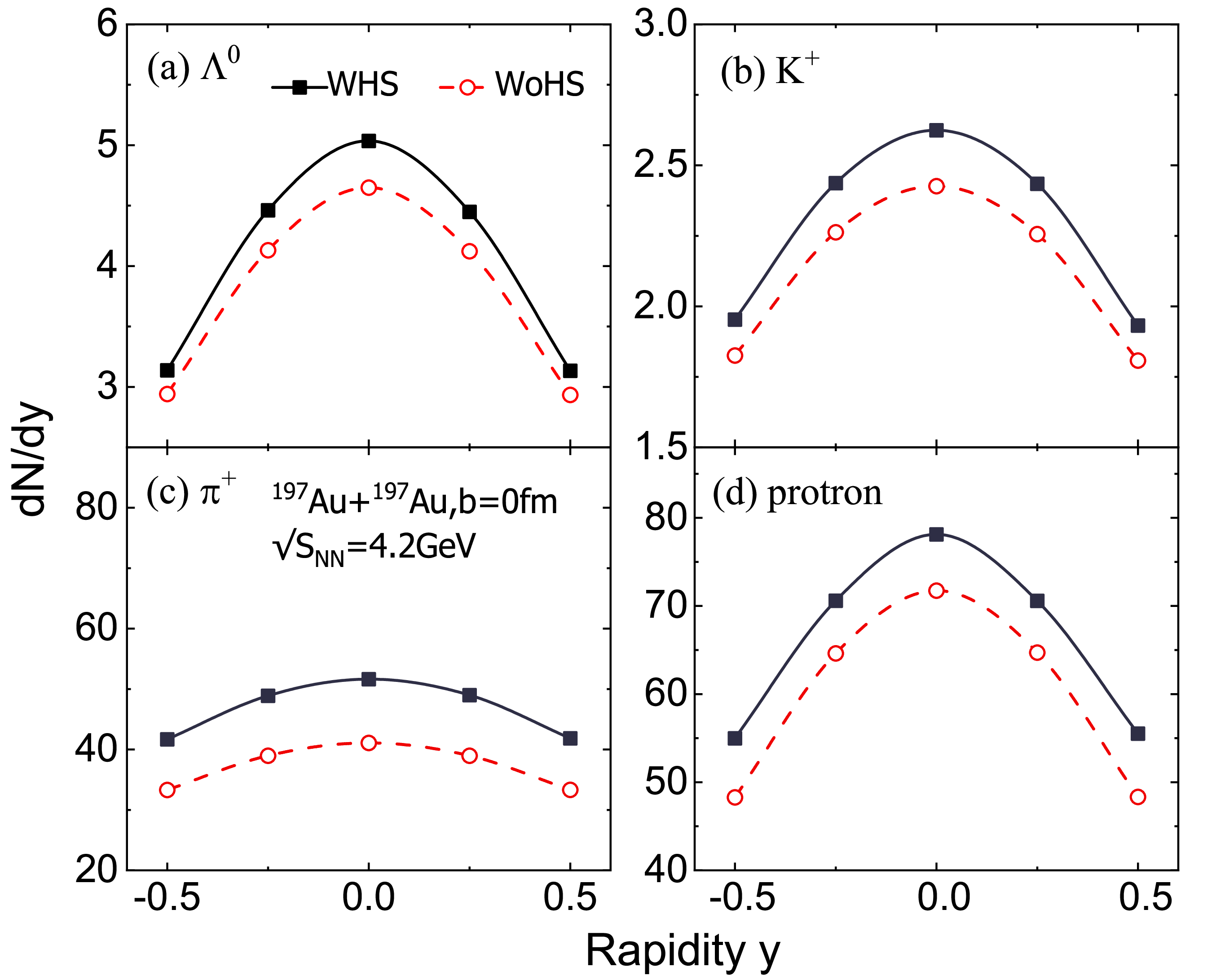}
\caption{Effects of hadronic rescatterings on the $\Lambda^0$, K$^+$, $\pi^+$, and proton productions in the central ${}^{197}\mathrm{Au}$+${}^{197}\mathrm{Au}$ collision at $\sqrt{s_{NN}}=4.2  \,\mathrm{GeV}$, given by the PACIAE model. The solid lines and filled squares (WHS) represent results with hadronic rescatterings while the dashed lines and open circles (WoHS) represent results without hadronic rescatterings.} \label{hadron}
\vspace{0.25cm}
\end{figure}
As described in the model introduction, PACIAE-B consists of two steps: the first step uses PYTHIA to simulate each hadron-hadron collision, and the second step handles hadron rescattering. To investigate the effect of hadron rescattering on particle yields, one can turn on and off the hadron rescattering part artificially. Figure~\ref{hadron} shows variations of the yields of different particles in the ${}^{197}\mathrm{Au}$+${}^{197}\mathrm{Au}$ collision before and after hadron rescatterings.
It can be clearly seen that hadronic rescatterings really increase the yields of different particles. Although these values differ numerically from the results given by the AMPT model as shown in Figure~\ref{caau}, the results provided by the PACIAE-B model here strongly support our previous conjecture that the increase in the particle productions in heavier reaction system is due to the hadronic rescatterings.

It should be noted that, currently, PACIAE-B or C have hardly been used to study nucleus-nucleus collisions below 10 GeV. Here, we utilize the PACIAE model solely to investigate the effects of partonic inelastic scatterings and hadronic rescatterings on particle production, in order to validate our earlier discussions and hypotheses. In recent years, the AMPT model in fact has been successfully applied to a large number of studies on nucleus-nucleus collisions below 10 GeV. Therefore, we do not conduct comparative studies of the models here.

%
In summary, based on the relativistic heavy-ion collisions' transport model, we studied the particle productions in heavy and light reaction systems $^{40}$Ca+$^{40}$Ca, $^{48}$Ca+$^{48}$Ca, and $^{197}$Au+$^{197}$Au at $\sqrt{s_{NN}}=4.2  \,\mathrm{GeV}$. It is found that the ratios of identical particles such as $\Lambda^0$, K$^+$, $\pi^+$, and proton produced in these systems are sensitive to whether quark matter is produced during the heavy-ion collisions. The primary reason is that, in pure hadronic transport, heavier reaction systems lead to more frequent multiple scatterings of hadrons compared to lighter ones. The increase in multiple scatterings is greater than the proportional increase in the number of participating nucleons, resulting in higher particle production. However, in scenarios where quark matter is present in heavy-ion collisions, the ratios of particle yields of the same kind in heavy and light reaction systems are significantly lower than those in purely hadronic transport. The reason is that multiple scatterings among partons do not lead to an increase in their numbers. Therefore, we propose that these particle yield ratios (especially the ratios for $\Lambda^0$ or K$^+$) between heavy and light reaction systems can serve as probes for detecting the hadron-quark phase transition in heavy-ion collisions. Furthermore, using the double-system particle yield ratio as a probe has the advantage of reducing systematic errors in both theoretical calculations and experimental measurements.

Experimental measurements from the E866/E917/E895 Collaborations \cite{exp7,exp8} in heavy-ion collisions at center-of-mass energies relevant to this study (e.g., $\sqrt{s_{NN}} \approx 4.2$ GeV) are crucial for validating the proposed sensitivity of particle emission to the hadron-quark phase transition. Systematic comparisons of particle yields between heavy collision systems (e.g., Au+Au) and light collision systems (e.g., Ca+Ca) can reveal deviations in strange particle production ratios, which depend on the presence or absence of quark matter. Such multi-system studies are essential for isolating phase-transition signatures from background hadronic effects. By supplementing the measurements from the E866/E917/E895 Collaborations \cite{exp7,exp8} with data from light reaction systems like Ca+Ca, it will be possible to determine whether quark matter is produced through comparisons with theoretical calculations.

%
The author G.C.Y. extends gratitude to Dr. A.-K. Lei for the valuable discussions regarding the utilization of the PACIAE model. This work is partly supported by the National Natural Science Foundation of China under Grant Nos. 12275322, 12335008 and CAS Project for Young Scientists in Basic Research YSBR-088.


\begin{thebibliography}{100}
\bibitem{ellis2006little}
J.~Ellis, J. Phys. Conf. Ser. {\bf 50}, (2006) 8.
\bibitem{iancu2012qcd}
H.~St\"{o}cker, W.~Greiner, Phys. Rep. {\bf 137}, (1986) 277.
\bibitem{odyniec2022probing}
G.~Odyniec, Probing the QCD phase diagram with heavy-ion collision experiments, in: D.~Blaschke, K.~Redlich, C.~Sasaki, L.~Turko (Eds.), Understanding the origin of matter: perspectives in quantum chromodynamics, Springer, 2022, pp. 3--29.
\bibitem{aidala2023new}
C.~Aidala, A.~Aprahamian, P.~Bedaque, L.~Bernstein, J.~Carlson, M.~Carpenter, K.~Chipps, V.~Cirigliano, I.~Cloet, A.~de~Gouvea \emph{et~al.}, A new era of discovery: The 2023 long-range plan for nuclear science, California, 2023.
\bibitem{heinz2001sps}
U.~Heinz, AIP Conf. Proc. {\bf 602}, (2001) 281.
\bibitem{gyulassy2004qgp}
M.~Gyulassy, The QGP discovered at RHIC, in: W.~Greiner, M. G.~Itkis, J.~Reinhardt, M. C.~G\"{u}\c{c}l\"{u} (Eds.), Structure and dynamics of elementary matter. Springer, Netherlands, 2004, pp. 159--182.
\bibitem{niida2021signatures}
T.~Niida, Y.~Miake, AAPPS Bull. {\bf 31}, (2021) 12.
\bibitem{schmidt2006lattice}
C.~Schmidt, Lattice QCD at finite density, in: T.~Blum, M.~Creutz, C.~DeTar, F.~Karsch, A.~Kronfeld, C.~Morningstar, D.~Richards, J.~Shigemitsu, D. Toussaint (Eds.), Part of Proceedings, 24th International Symposium on Lattice Field Theory (Lattice 2006), Tucson, USA, July 23-28, 2006. PoS LAT2006, 2006, pp. 021.
\bibitem{du2024qcd}
L.~Du, A.~Sorensen, M.~Stephanov, Int. J. Mod. Phys. E {\bf 33}, (2024) 2430008.
\bibitem{stephanov2006qcd}
M.~A.~Stephanov, in: T.~Blum, M.~Creutz, C.~DeTar, F.~Karsch, A.~Kronfeld, C.~Morningstar, D.~Richards, J.~Shigemitsu, D.~Toussaint (Eds.), 24th International Symposium on Lattice Field Theory (Lattice 2006) : Tucson, USA, July 23-28, 2006. PoS LAT2006, 2006, pp. 024.
\bibitem{csernai1999third}
L.~P. Csernai, D.~R{\"o}hrich, Phys. Lett. B {\bf 458}, (1999) 454.
\bibitem{shen2021studying}
C.~Shen, Nucl. Phys. A {\bf 1005}, (2021) 121788.
\bibitem{gajdovsova2021probing}
K.~K. Gajdo{\v{s}}ov{\'a}, Nucl. Phys. A {\bf 1005}, (2021) 121802.
\bibitem{baier2002jet}
R.~Baier, Nucl. Phys. A {\bf 715}, (2003) 209.
\bibitem{stephanov1999event}
M.~Stephanov, K.~Rajagopal, E.~Shuryak,  Phys. Rev. D {\bf 60}, (1999) 114028.
\bibitem{ASAKAWA2016299}
M.~Asakawa, M.~Kitazawa, Prog. Part. Nucl. Phys. {\bf 90}, (2016) 299.
\bibitem{shuryak2020light}
E.~Shuryak, J.~M. Torres-Rincon, Eur. Phys. J. A {\bf 56}, (2020) 241.
\bibitem{patra2009j}
J.~Qiu, J.~P. Vary, X.~Zhang, Phys. Rev. Lett. {\bf 88}, (2002) 232301.
\bibitem{dusling2008dilepton}
K.~Dusling, S.~Lin, Nucl. Phys. A {\bf 809}, (2008) 246.
\bibitem{YONG2023138051}
G. C.~Yong, Phys. Lett. B {\bf 843}, (2023) 138051.
\bibitem{hyd1}Yukinao Akamatsu \emph{et~al.}, Phys. Rev. C {\bf 98}, 024909 (2018).
\bibitem{hyd4}Jakub Cimerman, Iurii Karpenko, Boris Tom\'{a}\v{s}ik, and Pasi Huovinen, Phys. Rev. C {\bf 107}, 044902 (2023).
\bibitem{hyd3}Anna Sch\"{a}fer, Iurii Karpenko, Xiang-Yu Wu, Jan Hammelmann, Hannah Elfner, Eur. Phys. J. A {\bf 58}, 230 (2022).
\bibitem{hyd5}Lipei Du, Han Gao, Sangyong Jeon, and Charles Gale, Phys. Rev. C {\bf 109}, 014907 (2024).
\bibitem{hyd6}K. Werner, J. Jahan, I. Karpenko, T. Pierog, M. Stefaniak, and D. Vintache, Phys. Rev. C {\bf 111}, 014903 (2025).
\bibitem{hyd2}Chun Shen, and Sahr Alzhrani, Phys. Rev. C {\bf 102}, 014909 (2020).
\bibitem{lin2005multiphase}
Z. W.~Lin, C. M.~Ko, B. A.~Li, B.~Zhang, S.~Pal, Phys. Rev. C {\bf 72}, (2005) 064901.
\bibitem{yong2021double}
G. C.~Yong, Z. G.~Xiao, Y.~Gao, Z. W.~Lin, Phys. Lett. B {\bf 820}, (2021) 136521.
\bibitem{lei2023introduction}
A. K.~Lei, Y. L.~Yan, D. M.~Zhou, Z. L.~She, L.~Zheng, G. C.~Yong, X. M.~Li, G.~Chen, X.~Cai, and B. H.~Sa, Phys. Rev. C {\bf 108}, (2023) 064909.
\bibitem{yong2024pt}G. C.~Yong, Phys. Lett. B {\bf 848}, (2024) 138327.
\bibitem{amptrw}Z. W.~Lin, L.~Zheng, Nucl. Sci. Tech. {\bf 32}, (2021) 113.
\bibitem{exp7}L. Ahle \emph{et~al.} (E866/E917 Collaboration), Phys. Lett. B {\bf 490}, 53 (2000).
\bibitem{exp8}J. L. Klay \emph{et~al.} (E895 Collaboration), Phys. Rev. C {\bf 68}, 054905 (2003).
\end{thebibliography}
\end{document}